\documentclass[superscriptaddress,reprint,amsmath,amssymb,pra]{revtex4-1}

\usepackage{graphicx}% Include figure files
\usepackage{dcolumn}% Align table columns on decimal point
\usepackage{bm}% bold math
\usepackage{hyperref}% add hypertext capabilities
\usepackage{float}
\usepackage{ulem}
\usepackage{amsmath}
\usepackage[dvipsnames]{xcolor}

\begin{document}

\title{
Improving entanglement generation rates in trapped ion quantum networks using nondestructive photon measurement and storage\\
}

\author{John Hannegan}
\altaffiliation{These authors contributed equally to this work}
\author{James D. Siverns}
\altaffiliation{These authors contributed equally to this work}
\author{Jake Cassell}
\affiliation{IREAP and the University of Maryland, College Park, Maryland 20742, USA}
\author{Qudsia Quraishi}
\affiliation{Army Research Laboratory, 2800 Powder Mill Road, Adelphi, Maryland 20783, USA}
\affiliation{IREAP and the University of Maryland, College Park, Maryland 20742, USA}
\homepage{https://quraishi.physics.umd.edu}
\date{\today}

\begin{abstract}
Long range quantum information processing will require the integration of different technologies to form hybrid architectures combining the strengths of multiple quantum systems. 
In this work, we propose a hybrid networking architecture designed to improve entanglement rates in quantum networks based on trapped ions. Trapped ions are excellent candidates as network nodes but photon losses make long-distance networking difficult. 
To overcome some losses and extend the range of trapped-ion-based networks, we propose including neutral-atom-based non-destructive single-photon detection and single photon storage in between networking nodes, forming a hybrid network. 
This work builds on recently demonstrated optical frequency conversion of single photons emitted by trapped ions. 
We derive the average two-node entanglement rate for this proposed network architecture as a function of distance. Using reasonable experimental parameters, we show this proposed quantum network can generate remote entanglement rates up to a factor of 100 larger than that of an equivalent homogeneous network at both near-IR and C-band wavelengths for distances up to 50 km.

\end{abstract}

\maketitle

\section{\label{sec:Intro}Introduction}
Integrating different quantum networking elements together to form a hybrid network could offer a wider range of capabilities over those based purely on homogeneous components. 
Two well-developed quantum networking components with complimentary properties are trapped ion and neutral-atom systems.
Trapped ions are excellent matter qubits due to their ability to perform high-fidelity local quantum operations \cite{Harty2014,Ballance2016,Gaebler2016}, making them a leading technology for quantum computation \cite{Debnath2016,erhard2019,wright2019,pino2020}, simulation \cite{Smith2016,Zhang2017,Hempel2018,Jurcevic2017} and quantum networking nodes \cite{Stephenson2019,Hucul15}. 
However, trapped ion systems presently tend to have relatively low photon collection efficiencies and limited photonic propagation distances, reducing their effectiveness in long distance networking. 
Neutral-atom systems utilize large optical non-linearities and strong light-matter interactions \cite{Lampen18} making them well suited for high-efficiency photon production \cite{dudin2012}, storage \cite{eisaman2005,novikova2012,Leong20}, and non-destructive photon measurements \cite{reiserer2013,Xia2016}. A hybrid network combines the strengths of each system to overcome limitations present in homogeneous networks.

Entanglement between ions, as well as other matter qubits, has been demonstrated in homogeneous two-node networks relying on the interference of two photons, each entangled with their corresponding matter qubit \cite{simon2003,Moehring07,Stephenson2019,ritter2012,lettner2011,hensen2015,stockill2017,humphreys2018}.
Such networks typically rely on probabilistic entanglement heralding protocol where, after every photon request, each node must wait for a response from a Bell-state analyzer (BSA) before requesting another photon.
For the case of trapped-ion based nodes, the low photon collection efficiencies reported \cite{siverns2017ion} means that most entanglement attempts yield a null result because no photon has been collected.
This leads to long dead times as null events from the BSA measurement must be fed back to each node.
Increasing photon collection efficiency remains an on-going challenge with efforts ranging from the use of custom-designed cavities \cite{takahashi2020} and other in-vacuo optics \cite{araneda2020,shu2011} to ex-vacu custom multi-element lenses \cite{Stephenson2019,crocker2019}. 

In this work, we propose integrating neutral-atom-based nondestructive single photon measurement (NDSPM) and photonic storage into a trapped ion-based network to  increase ion-ion entanglement rates.
The NDSPM acts as a flag for the presence of an ion-produced photon, allowing for the request of a new photon based on the result of the flag rather than the result of the BSA measurement. This reduces much of the dead time arising from photon loss and increases the photon request rate of each node. 
Photonic storage can then be placed at the inputs of the BSA to ensure photon synchronization at the BSA, allowing for more efficient use of photons produced by each node. 
Quantum frequency conversion (QFC) is used to make the optical frequency of ion-produced photons compatible with neutral-atom based-technologies \cite{Siverns19convert,Siverns19slow,Craddock19} and to produce telecom-wavelength photons suitable for long-distance networking \cite{Walker2018,Krutyanskiy2019,Bock2018}. 
Using reasonable experimental parameters, we calculate relative entanglement rates at network distances of up to 50 km for Ba$^+$-based network nodes.
We show analytically and confirm via simulation that trapped ion nodes integrated with neutral-atom based systems can give improved two-node entanglement rates by over a factor of 100 as compared with an equivalent homogeneous network.

%\vspace*{5.95em}

\begin{figure*}[htbp]
\includegraphics[width=1\textwidth]{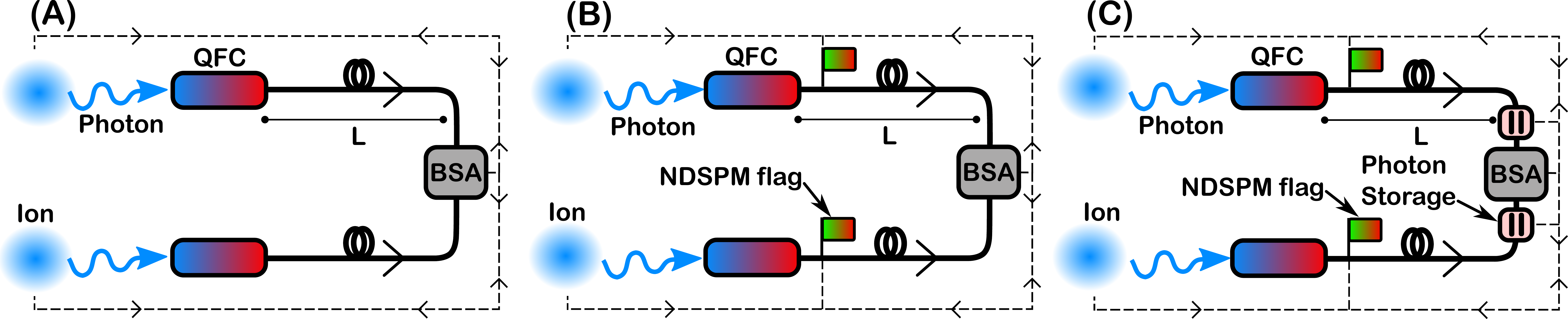}
\caption{\label{fig:Schematic} Layout of the proposed two-node networks. The nodes consist of trapped ions with QFC to provide optical frequency compatibility with NDSPM and photonic storage devices as well as create C-band photons for long distance transmission. (A) A standard homogeneous 2-node network as described in section \ref{Sec:Basic}. (B) and (C) show hybrid networks with a NDSPM flag placed directly after the conversion with additional photon storage devices are placed at the BSA in (C) as described in sections \ref{sec:NDSPM} and \ref{sec:Storage}. Classical signal channels are indicated with black dashed lines (depicted on lower network arm but present in both) with the fiber-based quantum channel indicated by the solid black lines. Extra QFC stages may be required before photonic storage and after NDSPM flags for C-band networks but are removed for clarity.}
\end{figure*}

%%%%%%%%%%%%%%%%%%%%%%%%%%%%%%%%%%%%%%%%%%%%%%%%%%%%5
\section{\label{sec:Components}Quantum networking with trapped ions and photons}
We consider a symmetric two-node network where each node contains a trapped ion capable of emitting a single photon, in the form of a flying qubit, entangled with its internal states \cite{Siverns17,Olmschenk2010,olmschenk2009}.
Photons may be requested from each node at a maximum rate, $r_{max}$, with the nodes synchronized to produce photons at the same time via a shared clock signal.
The ion-emitted photons are collected and coupled into at least one QFC setup connected to a network fiber of length $L$ to both reduce fiber loss and make their frequency compatible with NDSPM and photon storage devices (Fig.~\ref{fig:Schematic}).

In the following sections we will describe the entanglement generation rate between a pair of trapped ion nodes under three different configurations:
First, in section \ref{Sec:Basic}, we describe entanglement rates for a standard homogeneous ion-based network but incorporate QFC to extend the network range.
 Next, in section \ref{sec:NDSPM}, we describe entanglement rates when adding a neutral atom-based NDSPM to remove dead-time associated with waiting for a null-result to be returned from the BSA measurement.
Finally, in section \ref{sec:Storage} we describe the case where NDSPM and photonic storage are utilized to increase entanglement generation rates and improve the network efficiency.
%%%%%%%%%%%%%%%%%%%%%%%%%%%%%%%%%%%%%%%%%%%%%5
\subsection{Entanglement Rate, $R_{E}(L)$, for Homogeneous Two-Node Networks}\label{Sec:Basic}
Ion-ion entanglement via two-photon interference followed by a BSA measurement is typically  used to generate two-node entanglement \cite{Hucul15,Stephenson2019}. Such entanglement has been limited to networks with lengths, $L$, of only a few meters primarily due to photonic loss in fiber connections between nodes.
The two-node entanglement generation rate, $R(L)$, of these networks (Fig.~\ref{fig:Schematic}(A)) scale quadratically with detection probability (linearly for each node) as given by 

\begin{equation}
\label{eq:EntRate}
R(L) = (1/2)r(L)P^2_{B}(L),
\end{equation}

\noindent
where $r(L)$ is the synchronized photon request rate of the nodes and $P_{B}(L)$ is the probability of a photon being emitted, collected, coupled into a fiber and detected at the BSA. 
The factor of 1/2 accounts for half the BSA-detected signal yielding the desired entanglement \cite{calsamiglia2001}. The photon request rate,
$r(L)$, is given by the slower of $r_{max}$ or $(2t_{n}(L))^{-1}$, where $t_{n}(L)$ is the time taken for a photon to travel the length of the network via the quantum channel and the time for information to be fed back to the node from the BSA via the classical channel.
For most ion-based networks, the length of the network becomes the limiting factor after a few hundred meters.

For the network shown in Fig.~\ref{fig:Schematic}(A), $P_{B}(L)$ is given by

\begin{equation}
    P_{B}(L) = P_{p}P_{Q}^yP_{f}(L)P_{d}
\end{equation} \label{eq:PBSA}

\noindent
where $P_{Q}$ is the QFC efficiency, $P_{f}(L)$ is the probability of transmission along the network fiber, $P_{d}$ is the BSA detector efficiency, $P_p$ is the probability of a photon being emitted, collected and coupled into the network fiber and $y$ is the total number of QFC steps used per network arm. 
We assume the nodes are synchronized via a clock signal.
The probability of transmission along a fiber is modeled as an exponential loss as a function of the distance travelled along the fiber.

In practice, entanglement rates are greatly reduced by relatively low photon collection efficiencies (typically $\leq 10\%$ for trapped-ion systems \cite{Stephenson2019}).
This is because after each photon request event, network nodes experience a dead-time of $r(L)^{-1}$ waiting to receive the BSA measurement result to determine if entanglement was successful before proceeding to request another photon.

%%%%%%%%%%%%%%%%%%%%%%%%%%%%%%%%%%%%%%%%%%%%%%%%%%
\subsection{Entanglement Rate, $R'(L)$, with NDSPM}\label{sec:NDSPM}
The traditional two-node network, described in Section \ref{Sec:Basic} and shown in Fig.~\ref{fig:Schematic}(A), may be capable of a large photon request rate, $r(L)$, but in the presence of low photon coupling into the network, many of these attempts are wasted due to a low value of $P_p$. 
Placing a NDSPM device at both network nodes (Fig. \ref{fig:Schematic}(B)) which detects the presence of a travelling photon, without destroying the ion-photon entangled state, can serve as a herald to either allow the entanglement protocol to proceed or to trigger a node to request another photon.

When no photon is detected at the start of the network fiber, the NDSPM signal can be used to trigger a subsequent photon request at a time of $t_{nd}$, instead of $2t_{n}(L)$, where $t_{nd}$ is the NDSPM response time.
The response time of the NDSPM should not impede the entanglement protocol for a given network distance $L$, and so one requires $t_{nd}\ll 2t_{n}(L)$. 
In the example case analysis in Section \ref{sec:comparison} a NDSPM response time, $t_{nd}$, of 1 $\mu s$ is used.
Such rapid response is possible using a NDSPM scheme such as that proposed by Xia et al. \cite{Xia2016} which uses an effective three-wave mixing scheme in a neutral rubidium atomic vapor contained within a hollow-core photonic crystal fiber to impart a detectable phase shift on a probe beam when a single photon is present.
With this example, after measurement, photonic quantum state fidelities greater than 0.9 are expected, with detection efficiencies near 0.91 \cite{Xia2016}.
This fidelity is within the range where entanglement distillation could be used to purify the ion-ion entanglement produced in the network at the cost of entanglement rate \cite{nigmatullin2016}.
Alternatively, higher fidelities can be achieved by lowering the NDSPM detection efficiency for this method.
We note that other NDSPM implementations such as a neutral-atom coupled to an optical cavity  \cite{reiserer2013} and non-destructive single photon triggers \cite{Howell00} may be used to similar effect, provided they satisfy $t_{nd}\ll 2t_{n}(L)$ whilst sufficiently preserving the fidelity of the ion-photon entangled state.

The addition of a NDSPM flag makes it possible to request photons at a modified maximum rate of $r'_{max} = 1/T$, where $T = r^{-1}_{max} + t_{nd}$, until a photon is successfully detected in the network fiber.
The average photon request rate in this configuration, $r'(L)$, can be calculated using a weighted average between $2t_{n}(L)$ and $T$ given by

\begin{equation}\label{eq:Rdash}
    r'(L) =\frac{1}{p[2t_{n}(L)]+ (1-p)T},
    \end{equation}
    
\noindent
where the probability of a photon being detected, per-request by the NDSPM, is given by

\begin{equation}\label{eqn:p}
    p = P_p P^n_{Q} P_{nd},
\end{equation}

\noindent
where $P_{nd}$ is the NDSPM detection efficiency and $n$ is the number of QFC steps used before the NDSPM. In the example case in this work we consider a network with $n=1$.

Although Eq.~\ref{eq:Rdash} represents the average request rate of a single node, there is no guarantee that both nodes will be attempting to produce a photon at the same time. One node can successfully send a photon into the network, and be awaiting a signal from the BSA, while the other node is still attempting to produce a photon. 
The only instance in which entanglement can occur is when both nodes simultaneously attempt to produce photons (for a symmetric network) which occurs with probability

\begin{equation}\label{eq:alpha}
    \alpha(L) = \frac{\frac{1}{p}T}{\frac{1}{p}T+ 2t_{n}(L)}.
\end{equation}

This factor is equivalent to the ratio of the average time a node spends attempting to get a successful NDSPM, $T/p$, relative to the total time it takes on average to produce a photon and get a response from the BSA.

Analogous to Eq.~\ref{eq:EntRate}, the entanglement rate is then given by

\begin{equation}\label{eqn:RateDash}
    R'(L) = \alpha(L) r'(L)[P_{nd}\Gamma_{nd}P_{B}(L)]^2/2,
\end{equation}

where $\Gamma_{nd}$ represents the transmission of the NDSPM, allowing for the possibility of a photon being successfully detected, but not transmitted past the NDSPM device.
Such a transmission loss is not fundamental to the NDSPM scheme proposed in \cite{Xia2016}, but has been observed \cite{reiserer2013}.
Using experimental parameters presented in section \ref{sec:comparison}, we show that this rate can exceed that given by Eq.~\ref{eq:EntRate}.

%%%%%%%%%%%%%%%%%%%%%%%%%%%%%%%%%%%%%%%%%%
\begin{figure*}[!htbp]
\includegraphics[width=0.95\textwidth]{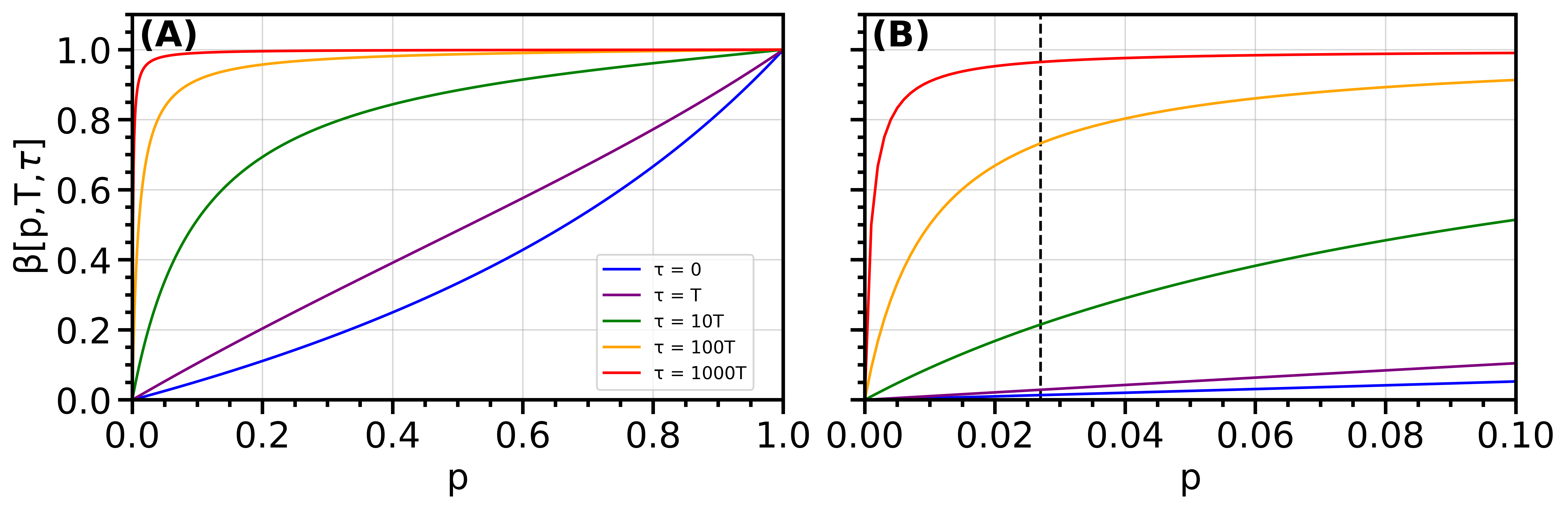}
\caption{(a) Multiplicative factor, $\beta(p,T,\tau)$, required to adjust entanglement rates as a function of $p$ for given ratios of storage time, $\tau$, to experimental rep rate, $T$. b) Zoom-in of $0 < p < 0.1$, which is a typical operating range of current trapped ion systems. The dashed vertical line indicates the value of $p$ used in Section \ref{sec:comparison}.\label{fig:beta}}
\end{figure*}
\subsection{Entanglement Rate, $R_{E}^{*}(L)$, with NDSPM and BSA Photonic Storage}\label{sec:Storage}
The network with a NDSPM described in section \ref{sec:NDSPM} can only produce entanglement when both nodes simultaneously attempt to produce photons, with probability $\alpha(L)$. 
We can remove the requirement for simultaneous photon emission on a given experimental cycle  by using photonic storage just before each input to the BSA to synchronize photon arrival times (Fig.~\ref{fig:Schematic}(C)).
A photon that successfully reaches the photonic storage at one BSA input is stored until the photonic storage for the other BSA input also contains a photon, whereupon both photons are released into the BSA.
This release event can be triggered by control electronics at the location of the BSA using logic circuitry and classical information fed forward by each NDSPM in the event of a successful photon herald. 
After the synchronized photon release from the storage elements, the result of the BSA measurement is fed back to the nodes so that they are again requested to produce photons.

This method will decrease the average attempt rate for each node, but with sufficient storage efficiency, will allow for more efficient use of photons produced by the nodes, and an increase in entanglement rate.
In this section, we will determine the average amount of time it takes for both nodes to produce a successful NDSPM, which is a requisite before entanglement may be attempted. 
We can then use this time to determine the entanglement rate of a hybrid network incorporating both NDSPM and photonic storage. 
The entanglement rate is compared with a homogeneous network's entanglement rate (Eq.~\ref{eq:EntRate}), for the case of a barium ion and neutral rubidium-based hybrid network in section \ref{sec:comparison}.

We begin by considering the number of photon-request attempts needed before node $i$ receives a successful NDSPM flag. The probability this occurs on the $x$th photon request since the last command from the BSA is given by a Geometric Distribution

\begin{equation}
    P_{i}(x) = (1-p)^{x-1}p,
\end{equation}

\noindent
where $i\in \{1,2\}$. 

The probability a node has successfully emitted a photon into the network after any of the first X repetitions is given by the cumulative distribution function,

\begin{eqnarray}
    \mathbb{P}_i(x \leq X) = 1-(1-p)^X.\label{eqn:prob1}
\end{eqnarray}

Assuming node 1 is successful, the two possible outcomes are that node 2 has either already produced a photon or that node 2 still needs to produce a photon.
In the former case, the production of a photon from node 1 is the limiting factor, with both nodes successfully sending a photon into the network by time $X T$.
For the latter case, node 2 is the limiting factor and entanglement can only be attempted after waiting an additional $1/p$ repetitions on average. The average time for both nodes to successfully send a photon into the network is then $X T + T/p$ for this case.
We can therefore calculate the average time between entanglement attempts as

\begin{eqnarray}\label{eqn:Tstar}
    T^{*}(L) = \sum_{X=1}^{\infty}P_{1}(X) \left[X  \mathbb{P}_2{(x \leq X)}\right.\nonumber\\
    \left.+ \left(X + p^{-1} \right)(1-\mathbb{P}_2{(x \leq X))} \right]T + 2t_{n}(L),
\end{eqnarray}

\noindent
where the additional term $2t_{n}(L)$ is the network round trip time described in Section \ref{Sec:Basic}.
Inserting Eq. \ref{eqn:prob1} into \ref{eqn:Tstar} and simplifying gives

\begin{equation}
T^{*}(L) = \frac{2p-3}{p(p-2)}T + 2t_{n}(L).
\end{equation}

The inverse of  $T^{*}(L)$ is then the node's effective repetition rate, $r^{*}(L)$.
We can write the entanglement generation rate in a manner similar to Eq.~\ref{eqn:RateDash} as 

\begin{equation}\label{eqn:RateStar}
R^{*}(L) = r^{*}(L)[P_{f}(L)\Gamma_{nd}E_{s}P^z_{Q}P_{d}]^2/2, 
\end{equation}

\noindent
where $E_{s}$ is the efficiency of the photon storage just before the BSA and where $z$ is the number of QFC steps used in each network arm after the NDSPM. (This is distinct from the total number of QFC steps, $y$, as defined in section \ref{Sec:Basic}).
Depending upon the color for used for network transmission and color needed for photon storage, there might be multiple QFC steps after the NDSPM involved as discussed in Section \ref{sec:comparison}.

\begin{figure*}[htbp!]
\includegraphics[width=0.95\textwidth]{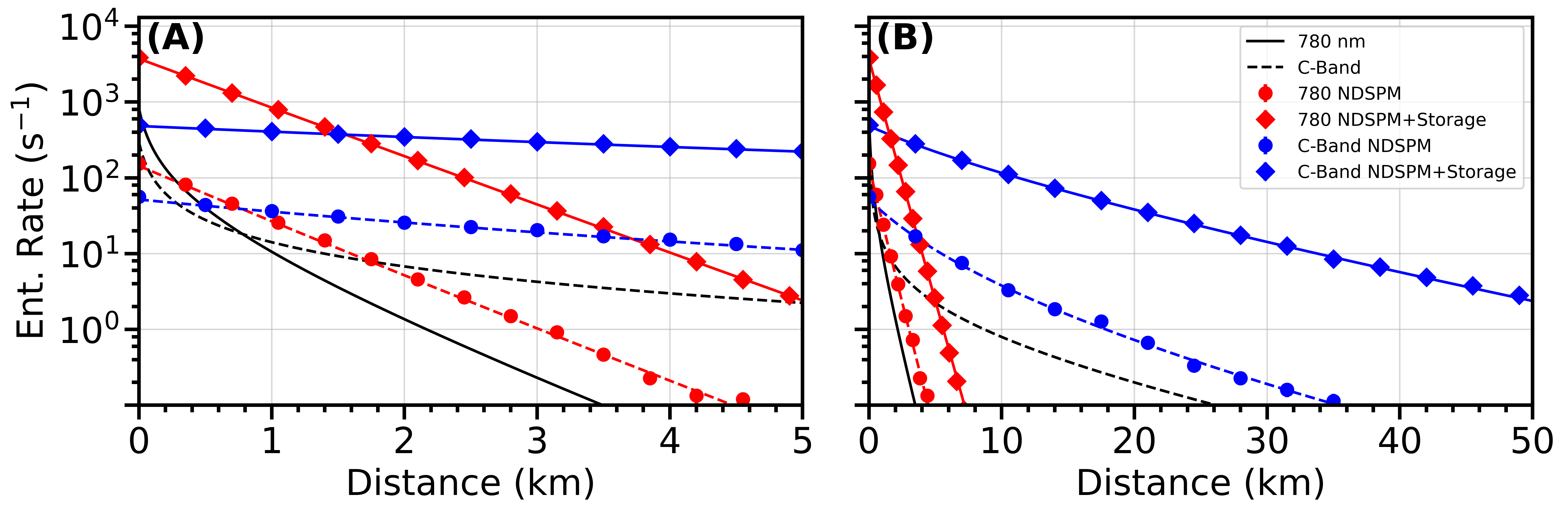}
\caption{\label{fig:EntRate}The entanglement generation rate of a two node network connected using 780 nm fiber links in a standard homogeneous network described in section \ref{Sec:Basic} (black solid), using NDSPM as described in section \ref{sec:NDSPM} (red dashed and circles) and using NDSPM and photon storage as described in section \ref{sec:Storage} (red solid and diamonds) as a function of network distance, $L$. Rates are also shown for a C-band-linked network in a standard homogeneous configuration (black dashed), using NDSPM (blue dashed and circles) and using NDSPM and photon storage (blue solid and diamonds). (A) shows the rates over 5 km while (B) shows them over 50 km. The theory curves are determined from the analytical solutions in the paper and the points are from simulated data with statistical error bars smaller than the plot points.}
\end{figure*}
\begin{figure*}[t]
\includegraphics[width=0.95\textwidth]{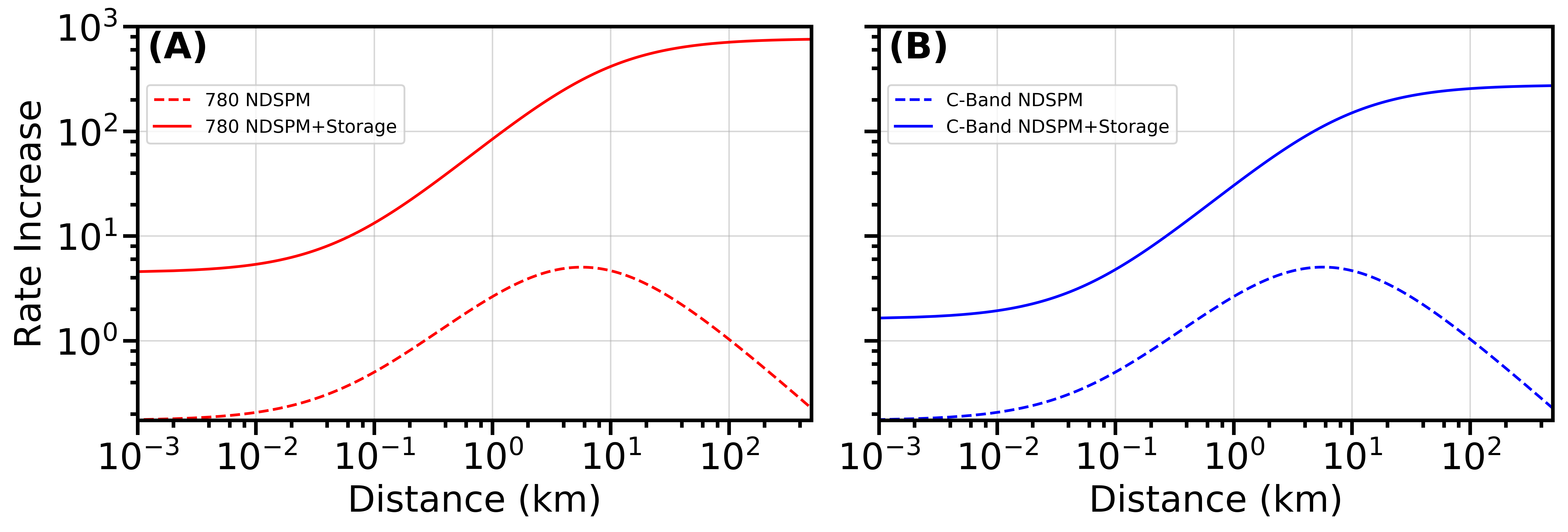}
\caption{\label{fig:Ratios}Entanglement rate increases are shown for networks with NDSPM (dashed) and NDSPM with storage (solid) over standard homogeneous networks using 780 nm photons, (A), and C-band photons, (B), as a function of distance, $L$.
These are calculated from the ratios of the relevant curves in Fig. \ref{fig:EntRate}.
In all cases the theory curves are determined from the analytical solutions in the paper.}
\end{figure*}
Any realistic storage will not be able to store photons indefinitely.
Assuming storage that decays exponentially in time with time constant $\tau$, one can show that $R^{*}(L)$ is modified by the multiplicative factor (see Appendix \ref{sec:NonInfiniteStorage})
\begin{equation}
   \beta(p,T,\tau) =\frac{p(1+e^{T/\tau}-p)}{(2-p)(e^{T/\tau}+p-1)}.
\end{equation}

Critically, this factor does not depend on the length of the network, and instead displays an exponential dependence on the ratio $T/\tau$. 
This factor is plotted as a function of $p$ for various ratios of $T/\tau$ in Fig. \ref{fig:beta}(a).
As current trapped ion systems have $p<0.1$, a zoom-in of the region $0<p<0.1$ is shown in Fig. \ref{fig:beta}(b).
For typical ion systems, $\tau \gtrsim 1000T$ will ensure that finite storage reduces entanglement rates by $<10\%$.
%%%%%%%%%%%%%%%%%%%%%%%%%%%%%%%%%%%%%%%%%%%%%%%%%
\section{Case Study Using Barium Ions and Rubidium Atoms}\label{sec:comparison}
To highlight the entanglement rate increase using the hybrid network tools described in this work, we present an example using network nodes comprised of single trapped Ba$^+$ ions.
The ions are optically excited with near-unity probability leading to the probabilistic emission of a single 493-nm photon which serves as a flying qubit maximally entangled with the ion's internal states \cite{Siverns17}.
We use a maximum node photon request rate of $r_{max} = 2$ MHz as a realistic maximum for current ion trap experiments \cite{Stephenson2019}.

\begin{figure*}[htbp!]
    \centering
    \includegraphics[width=0.95\textwidth]{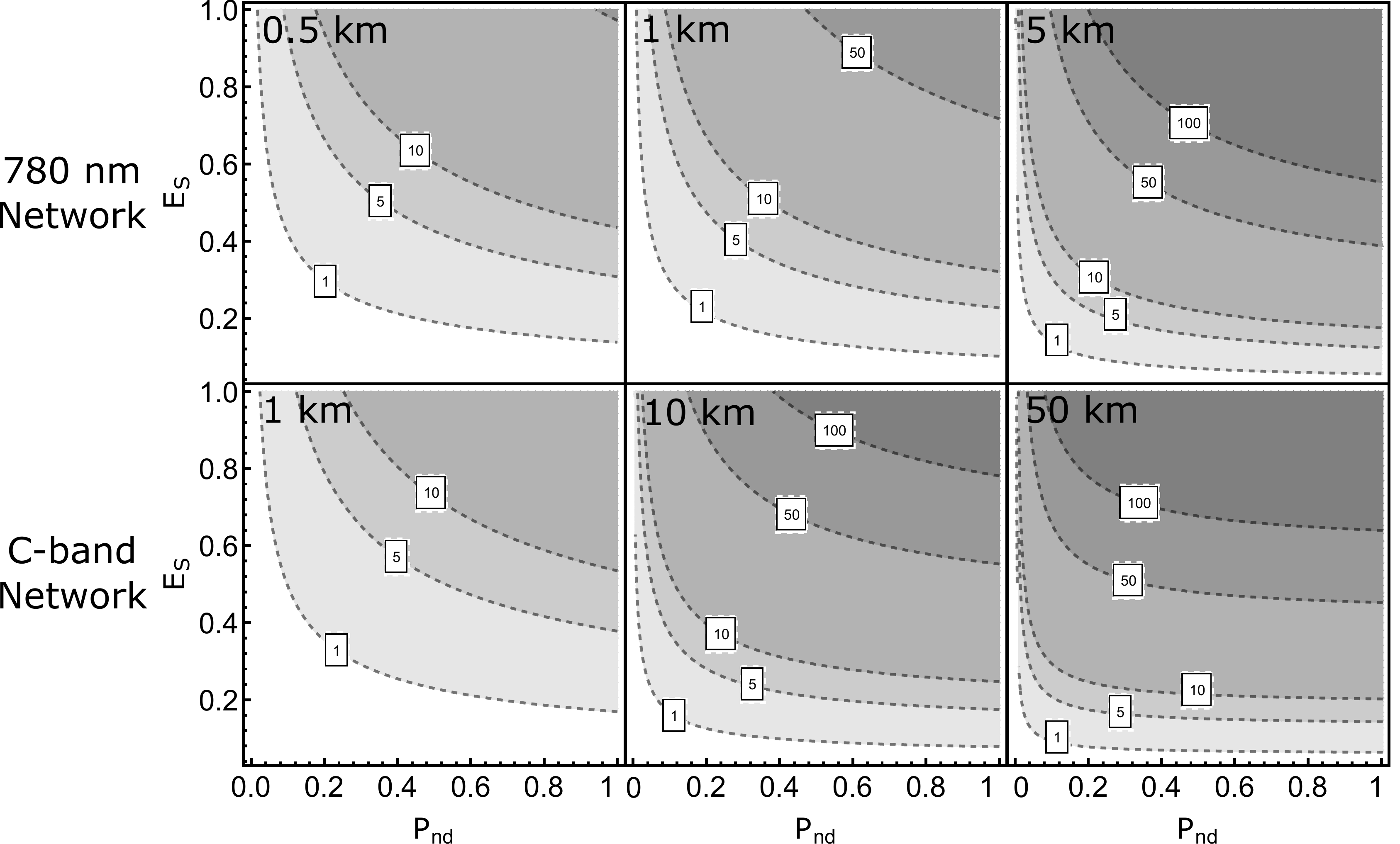}
    \caption{Entanglement rate increase contours at both 780 nm and C-band as a function of $E_s$ and $P_{nd}$ for various distances. All other variables are kept constant as given by Table \ref{tbl:variables} in Appendix \ref{Sec:Appendix:Table} and as mentioned in Section \ref{sec:comparison}.}
    \label{fig:contours}
\end{figure*}
We set the probability of a photon being emitted, collected and coupled into the network fiber as $P_p=0.06$ which assumes a 0.6 NA collection optic, 80$\%$ fiber coupling and an even Ba$^+$ isotope \cite{Siverns17}.
We use one QFC step ($y=1$) for converting the 493 nm photon to 780 nm, making it compatible with neutral Rb NDSPM devices (efficiency $P_{nd}=0.75$ corresponding to a fidelity of $>0.965$ using the method in \cite{Xia2016} and set $\Gamma_{nd} = 1$) and an additional QFC step ($y=2$) to take the photon frequency to C-band. 
We note that a three-step distillation process as outlined in \cite{nigmatullin2016} is projected to increase a fidelity of $>0.965$ to $\approx$ 0.9975 with an entanglement rate penalty of $\approx$ 8.
An additional QFC step is required for the C-band network using photonic storage (Section \ref{sec:Storage}) to take C-band back to 780 nm for compatibility with neutral-Rb storage devices (efficiency $E_s=1$).
Thus, $z=0$ and $z=2$ for the 780-nm and C-band based networks using storage, respectively.
The QFC efficiencies are set at 60$\%$ (see Appendix \ref{Sec:Appendix:QFC}).
We assume all network links comprise of optical fiber with a refractive index of 1.4 which is a good approximation for both near-IR and C-band networks. The fiber attenuation is set to 3 dB/km and 0.15 dB/km for the 780 nm and C-band network respectively.
Appendix \ref{Sec:Appendix:Table} gives the values used for entanglement rate calculations.

In Figs.~\ref{fig:EntRate} and \ref{fig:Ratios} we compare the entanglement generation rates for this example case in network configurations described in sections \ref{Sec:Basic}, \ref{sec:NDSPM} and \ref{sec:Storage} and operating at either 780 nm or in the C-band. 
The entanglement rates in Figs. \ref{fig:EntRate} and \ref{fig:Ratios} are determined using Eqs.~\ref{eq:EntRate}, \ref{eqn:RateDash} and \ref{eqn:RateStar} (lines) and also using simulated data (data points in Fig. \ref{fig:EntRate}). See Appendix \ref{Sec:Appendix:Simulations} for more details on the simulation.

The black curves in Fig.~\ref{fig:EntRate} show the entanglement generation rates for basic two-node networks at 780 nm (black solid) and C-band (black dashed) as described in section \ref{Sec:Basic}.
The dashed curves and circles show the rates for a 780 nm (red) and C-band (blue) network using NDSPM as described in section \ref{sec:NDSPM}. 
When compared to hybrid networks using NDSPM (circles) and using NDSPM and photonic storage (diamonds), as described in sections \ref{sec:NDSPM} and \ref{sec:Storage} respectively, there is an increase in entanglement rates.
For example, the 780 nm network with NDSPM outperforms a basic network using C-band photons up to $\approx$ 1.8 km. 
By adding the photonic storage this distance can be increased to $\approx$5 km (red diamonds in Fig.~\ref{fig:EntRate}). 
Similarly, Fig. \ref{fig:EntRate}(A) shows that a C-band network using NDSPM outperforms the 780 nm network using NDSPM and photonic storage after $\approx$ 3.8 km.

The ratio of entanglement rate between a hybrid network using NDSPM(Eq. \ref{eqn:RateDash}) and a homogeneous network (Eq. \ref{eq:EntRate}) shows a peak followed by a slow decline (dashed curves in Fig. \ref{fig:Ratios}). 
This decline can be explained by Eqs.~\ref{eq:Rdash} and \ref{eq:alpha} where $2t_n(L)$ will dominate at large distances and, eventually, this method \textit{underperforms} the homogeneous network rate $R(L)$. 
Comparing Eqs. \ref{eq:EntRate} and \ref{eqn:RateDash}, this occurs at lengths such that $P_{nd}^2 \frac{T}{2 p^2} \lesssim t_n(L)$.
The 780-nm network using photonic storage in addition to NDSPM (Eqn. \ref{eqn:RateStar}) for each node outperforms a standard 780-nm network at all distances for the values used in this work (solid curves in Fig. \ref{fig:Ratios}). 
At long distances $T^*(L) \approx r(L)^{-1}$, and for the 780-nm network, the rate ratio approaches an asymptotic limit given by $[\Gamma_{nd}E_{s}/(P_{p}P_{Q})]^{2} \approx 772$, the increase in success probability given a successful NDSPM.
The C-band network with storage, having an additional stage of QFC versus the non-storage case, approaches and asymptotic limit of $[\Gamma_{nd}E_{s}/P_{p}]^{2} \approx 278$.

The example cases in Figs.~\ref{fig:EntRate} and \ref{fig:Ratios} use set values of the photon storage efficiency, $E_s = 1$, and NDSPM efficiency, $P_{nd} = 0.75$. 
Variations in these values will affect the network’s entanglement rate increase over a homogeneous network.
In Fig.~ \ref{fig:contours} we show contour plots for both the 780 nm and C-band based networks as a function of $E_s$ and $P_{nd}$.
In both the 780 nm and C-band networks it can be seen that entanglement generation rates over an order of magnitude compared with a homogeneous network can be achieved with only modest values of $E_s$ and $P_{nd}$. In fact, for the C-band network, factors of over 100 can be achieved after 10 km with $E_s$ and $P_{nd}$ values of $\approx$ 0.6.

We have described how integration of neutral-atom-based technologies into a trapped-ion based quantum network can overcome photon losses to yield significant increases in entanglement generation rates.
We show this increase can be over a factor of 100 in both 780 nm and C-band based fiber network links.
The use of hybrid technology in trapped-ion based quantum networks is a promising method for establishing quantum networks with projected gains over their homogeneous counterparts in entanglement generation rates.

\begin{acknowledgments}
All authors acknowledge support from the United States Army Research Lab's Center for Distributed Quantum Information (CDQI) at the University of Maryland and the Army Research Lab.
\end{acknowledgments}

\appendix
\section{QFC Efficiencies}\label{Sec:Appendix:QFC}
We have performed frequency conversion efficiency tests using a waveguide buried in a periodically-poled LiNbO$_3$ (PPLN) crystal fabricated via reverse proton exchange \cite{Srico}. 
Within the waveguide, difference frequency generation results in the conversion of 493-nm photons to 780-nm photons in the presence of a high intensity pump \cite{kumar1990,Siverns17,Siverns19convert}.
A high intensity pump laser operating at 1343 nm as well as low intensity 493-nm light are combined via a dichroic mirror and then coupled into one side of the waveguide using an uncoated aspheric lens.
At the output of the PPLN waveguide, another uncoated aspheric lens is used to roughly collimate the converted 780-nm light.
The converted signal is then separated from the pump through use of both a dichroic mirror as well as a set of band-pass (Semrock 1326/SP-25 and Semrock LL01-780-12.5) filters and coupled into a polarization maintaining single mode fiber.
The power of the converted signal is then measured at the output of this short fiber using a standard power meter. 
This is similar to the setups used in \cite{Hannegan2021}.

\begin{figure}[!htbp]
\includegraphics[width=0.45\textwidth]{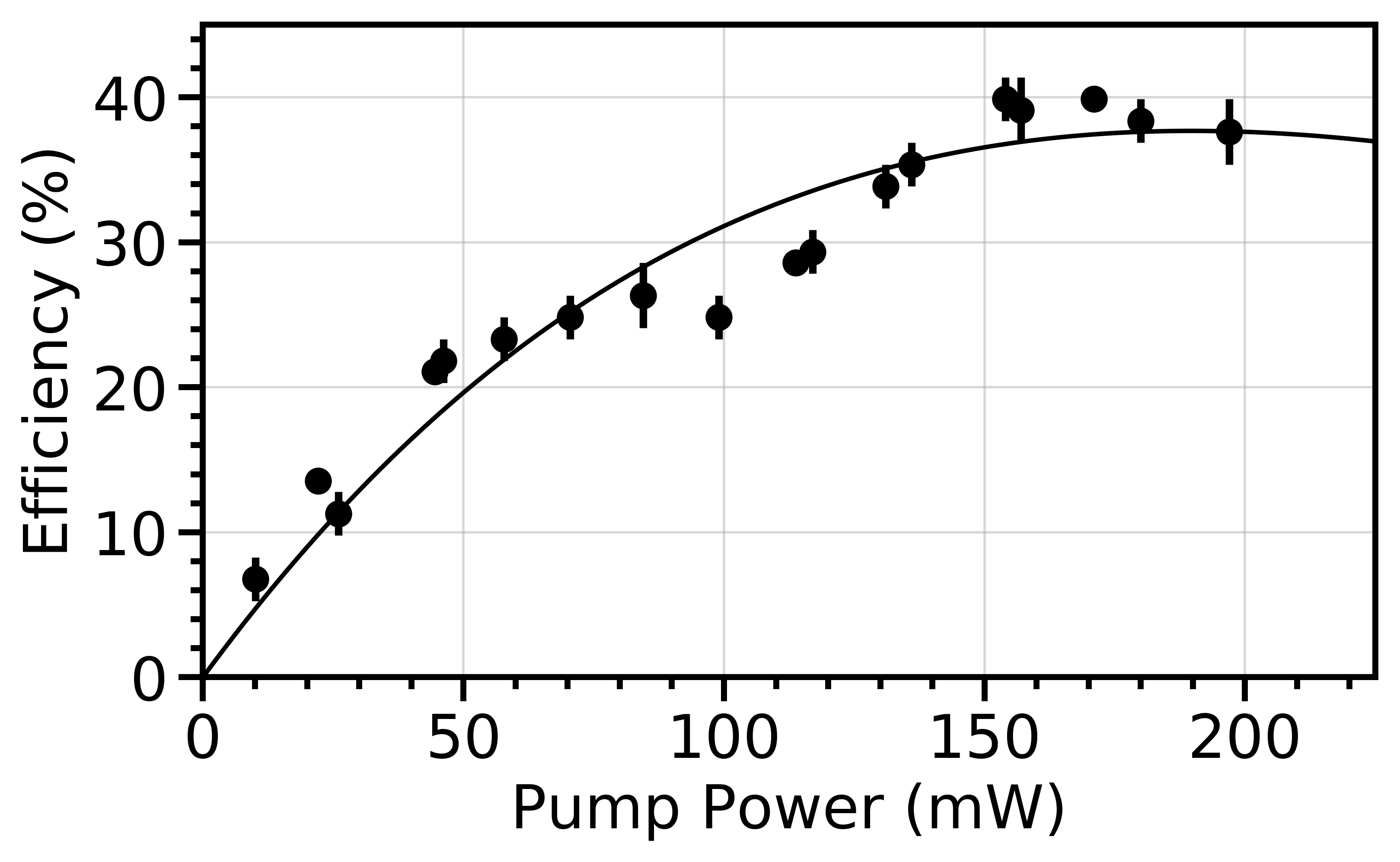}
\caption{\label{fig:QFCcurves} Efficiency of QFC from 493 nm to 780 nm as a function of pump power, $P$, coupled into a LiNbO$_3$ waveguide \cite{Srico}. The solid line is a fit to $\eta sin^2[(\pi/2)\sqrt{P/P_{m}}]$ where $P_{m}$ is the power at maximum conversion efficiency and $\eta$ is the maximum conversion efficiency.
Deviations of the data from theory fit are thought to be due to pump-induced phase-matching changes for this particular waveguide. 
Uncertainties result from converted signal power fluctuation over the course of the measurement.}
\end{figure}

We observe end-to-end QFC efficiencies of $\approx 40\%$ in conversion of 493-nm photons to 780-nm photons, including coupling back into optical fiber (Fig.~\ref{fig:QFCcurves}). 
An end-to-end efficiency $\sim 60\%$ could be achieved via the incorporation of anti-reflection coatings on coupling optics and on the facets of the LiNbO$_3$ waveguide \cite{Srico}, neither of which were used in the taking of the data in Fig.~\ref{fig:QFCcurves}.
We, therefore, assume $60\%$ QFC efficiency from 493 nm to 780 nm for this work and assume the same efficiency for QFC from 780 nm to C-band wavelengths, as has been previously demonstrated \cite{van2020}.
The output frequency of the QFC setup may be tuned into or out of optical resonance with neutral-rubidium systems (or to match other systems) by tuning the QFC setup's pump laser frequency.

\section{Network Simulations}\label{Sec:Appendix:Simulations}
As verification of the theory presented in the main text, we run a simulated experiment. 
The simulation iterates over a set number of clock cycles, with a cycle period equal to $T$ as defined in section \ref{sec:NDSPM}. 
All photon-emission attempts are implemented at an integer number of clock-cycles, as would be needed experimentally for synchronization purposes, and all travel times are rounded up to the nearest number of cycles for ease of coding. 
Events that can occur in the experiment with non-unity probability are simulated via comparison to random number generators (RNGs) with a precision of $10^{-4}$.
A Boolean flag variable is used for each channel, which can be set to True, after a successful photon detection event, or False, when no photon is detected or when the experiment should be reset.

First, we discuss the configuration presented in Fig.~\ref{fig:Schematic}(b) described in Section \ref{sec:NDSPM}.  
At every clock cycle,  equivalent to one iteration of the program loop, the program checks the flag variable for each channel, and if a flag is not raised (i.e. it is False), a photon generation attempt is performed using a comparison between an RNG and the per-shot probability a photon is detected by the NDSPM, given by Eq.~\ref{eqn:p} in the main text.
In the case of a successful NDSPM, the corresponding flag variable is set to True, preventing photon production from that channel for the amount of time it would take for information to travel to the BSA and back, $2t_{n}(L)$, rounded to the nearest number of clock cycles, after which the flag is reset to False.
In the case of a successful NDSPM, on the same cycle, the probability of a photon making it to the BSA is compared to an RNG to determine if the photon makes it through the network and if so, a Boolean variable for the BSA is set to True.
If both channels have a BSA variable set to True an entanglement event is recorded. The error in the entanglement rates are determined from statistical errors after $10^8$ program iterations.
At the end of every program iteration, the BSA variable for each channel is set to False to ensure that entanglement events are only recorded when both nodes successfully produce a photon on the same clock cycle.

We now discuss the simulation for the network shown in Fig.~\ref{fig:Schematic}(c) and discussed in section \ref{sec:Storage}, which removes the requirement that both photons be produced on the same clock cycle. 
In this case, the flag variables for each channel are set to True in the same manner as above, but are \textit{not} reset to False after a set time.
Instead, a successful NDSPM results in a storage variable for that channel being set to True after a time $t_{n}(L)$, the time it takes for a photon to travel from each node to the BSA.
After the NDSPM, BSA variables for each channel may be set to True in the same manner as discussed for the non-storage case, with the addition inclusion of storage efficiency in the photon transmission probability.
In this case however, BSA variables are \textit{not} automatically set to False at the beginning of every cycle, as photons can now be stored at the BSA. This allows one channel to preserve its photons in the storage element while the other channel continues photon-generation attempts.

Every cycle, the simulation checks to see if the storage variables for both channels are simultaneously set to True. 
At this point, if the photon variables for both channels are set to True, an entanglement event is recorded.
Both photon variables are then set to False, and after a time $t_{n}(L)$, to allow for information to travel back to each node, the flag variables for each channel are set to zero to allow photon production to recommence. 

The code for this simulation is available at (\url{https://github.com/ionquantumnetworks/TrappedIonPhotonMeasStorage}) or on request.

\section{Effect of non-infinite storage time}\label{sec:NonInfiniteStorage}

The entanglement rates given in the main text assume that the photons from each node can be stored indefinitely. 
Any realistic storage element will, however, have limits on the storage time.
We will examine the effect of this finite storage time on entanglement rates.

When a photon from one of the nodes reaches the storage element, it is stored until the other node's photon is stored. 
Critically, this means that the relevant time scale is the difference between the arrival times of each of the photons to their respective storage devices.
Because the network length is symmetric the relative arrival time is equivalent to the relative time between successful NDSPM of photons produced by each node.
The travel time (and therefore the network length) has no affect on this time for the symmetric network considered.
The probability that the second node successfully produces a photon M cycles after the first node successfully produces a photon is given by

\begin{align}
    P_{12}(M)&=\sum_{N=1}^{\infty}P_1(N)P_2(N+M)\nonumber\\
    %&=\sum_{N=1}^{\infty}p(1-p)^{N-1} p(1-p)^{N+M-1}\\
    &=\sum_{N=1}^{\infty} p^2 (1-p)^{2N + M - 2}\nonumber\\
    &= \frac{p(1-p)^M}{2-p}.
\end{align}

Similarly, the probability that the second node produces a photon M cycles before the first node successfully produces a photon is given by

\begin{align}
P_{21}(M)&=\sum_{N=1}^{\infty}P_1(N)P_2(N-M)\theta(N-M)\nonumber\\
%&=\sum_{N=1}^{\infty}p(1-p)^{N-1} p(1-p)^{N-M-1}\theta(N-M)\\
&=\sum_{N=1}^{\infty}p^2(1-p)^{2N-M-2}\theta(N-M)\nonumber\\
&=\frac{(1-p)^{M}}{2-p}
\end{align}

\noindent
where $\theta(x)$ represents the left-side continuous Heaviside step function.

Using these values, we can calculate the weighted probability that the photons from both nodes are successfully stored.
We consider single photon storage that decays exponentially over time, with decay constant $\tau$. 
Both photons are then stored and successfully released with a probability, relative to the maximum storage probability of $E_s$, of

\begin{align}
     \beta(p,T,\tau) &= \sum_{M=0}^{\infty}[P_{12}(M) + P_{21}(M)] e^{-M T/\tau} - P_{12}(0) \nonumber\\
    %=\sum_{M=0}^{\infty}[\frac{p(1-p)^M}{2-p}+&\frac{p(1-p)^{-M}(p-1)^{2M}}{2-p})] e^{-M T/\tau}-\frac{p}{2-p}\\
    &=\frac{p(1+e^{T/\tau}-p)}{(2-p)(e^{T/\tau}+p-1)},
\end{align}
where T is the period as defined in Section \ref{sec:NDSPM}. 
The $P_{12}(0)$ term is subtracted to keep from double counting for $M=0$, where one should note $P_{12}(0)=P_{21}(0)$.
The entanglement rate then is altered to
\begin{equation}\label{eqn:RateStarBeta}
R^{*}(L) = \beta(p,T,\tau)r^{*}(L)[P_{f}(L)E_{s}P^z_{Q}P_{d}]^2/2. 
\end{equation}

\section{Values Used for Entanglement Rate Calculations}\label{Sec:Appendix:Table}
\begin{table}[!htb]
    \centering
    \begin{tabular}{c|c|c}
        \hline
        Variable & Description & Value Used\\
        \hline
        $r_{max}$ & maximum photon production rate & 2 MHz\\
        \hline
        $t_{nd}$ & NDSPM time & $1 \mu$s\\
        \hline
        $t_{n}(L)$ & network travel time & $1.4 L /c$\\
        \hline
        $P_p$ & per-shot photon collection efficiency & 0.06\\
        \hline
        $P_Q$ & QFC efficiency per stage & 0.60\\
        \hline
        $P_d$ & BSA detector efficiency & 0.8\\
        \hline
        $P_{nd}$ & NDSPM efficiency & 0.75\\
        \hline
        $E_{s}$ & Photon Storage Efficiency & 1\\
        \hline
        $\Gamma_{nd}$ & NDSPM transmission & 1\\
        \hline
    \end{tabular}
    \caption{Values used in entanglement rate calculations.}
    \label{tbl:variables}
\end{table}

\bibliography{hybrid}

\end{document}